\documentclass[10pt]{article}

\usepackage[round]{natbib}
\usepackage{fullpage}
\usepackage{setspace}
\usepackage{parskip}
\usepackage{titlesec}
\usepackage[section]{placeins}
\usepackage{xcolor}
\usepackage{breakcites}
\usepackage{lineno}
\usepackage{hyphenat}
\usepackage{algorithm}
\usepackage{algorithmic}
\usepackage{graphicx}
\usepackage[space]{grffile}
\usepackage{latexsym}
\usepackage{textcomp}
\usepackage{longtable}
\usepackage{tabulary}
\usepackage{booktabs,array,multirow}
\usepackage{amsfonts,amsmath,amssymb}
\usepackage[utf8]{inputenc}
\usepackage[english]{babel}
\usepackage{float}
\usepackage[bottom, hang]{footmisc} % Footnotes at the bottom, no indent
\usepackage{authblk}

\PassOptionsToPackage{hyphens}{url}
\usepackage[colorlinks = true,
            linkcolor = blue,
            urlcolor  = blue,
            citecolor = blue,
            anchorcolor = blue]{hyperref}

\renewenvironment{abstract}
  {{\bfseries\noindent{\abstractname}\par\nobreak}}
  {\bigskip}

\title{USE-LFA: A Data-Driven Framework for Urban Air Mobility Site Evaluation using Latent Factor Analysis}

\author[a]{Sungmin Sohn}
\author[a]{Namwoo Kim}
\author[b]{Mark Hansen}
\author[a,\footnote{Corresponding author (yoonjin@kaist.ac.kr). 
\\ \hspace*{1em}Email addresses: smsohn1997@kaist.ac.kr (S. Sohn), ih736x@kaist.ac.kr (N. Kim), mhansen@ce.berkeley.edu (M. Hansen).}]{Yoonjin Yoon}

\affil[a]{Department of Civil and Environmental Engineering, Korea Advanced Institute of Science and Technology, Daejeon, South Korea}
\affil[b]{Department of Civil and Environmental Engineering, University of California, Berkeley, CA, 94720, United States}

\date{}

\begin{document}

% Title
\maketitle

\begingroup
\let\center\flushleft
\let\endcenter\endflushleft
\endgroup

\selectlanguage{english}
\begin{abstract}

Urban air mobility (UAM) introduces new challenges for infrastructure planning, requiring data-driven approaches for sustainable site selection. This study proposes USE-LFA (Urban Site Evaluation using Latent Factor Analysis), a framework designed to support equitable and environmentally conscious siting of urban ports. Applying latent factor analysis to 25 urban attributes in Seoul, the framework identifies six latent factors, grouped into two dimensions: Suitability and Attractiveness. These dimensions are combined through a tunable prioritization metric, enabling alignment with local strategic goals. The analysis uncovers spatial typologies and clustered siting patterns, highlighting regional disparities in site potential. Sensitivity analysis demonstrates that small adjustments in the Suitability–Attractiveness weighting substantially affect viable site candidates, emphasizing the need for calibrated decision-making. USE-LFA facilitates interpretable and transferable analysis across different urban contexts and datasets, offering a scalable approach to integrating UAM and other emerging mobility systems into urban environments, while advancing sustainable and inclusive transport infrastructure development.

\end{abstract}%
\sloppy
\doublespacing

\section{Introduction}

The advent of Urban Air Mobility (UAM) signifies a paradigm shift in urban transportation, offering promising applications such as air taxis, air ambulances, and air shuttles \citep{easa2021study, goyal2018urban}. UAM holds great potential to reduce congestion, improve emergency response times, and alleviate environmental pollution \citep{yedavalli2019assessment}. Ensuring its sustainability requires not only environmentally efficient aircraft designs that minimize emissions and noise pollution \citep{afonso2021design} but also the strategic placement of vertiports\(-\)facilities where UAM aircraft can takeoff, land, and be serviced. The vertiport site selection problem addresses this multifaceted challenge that requires comprehensive consideration of various urban factors to ensure effective, efficient, and sustainable integration of UAM systems into urban environments \citep{holden2016fast, cohen2021urban, garrow2021urban, yoon2025integrating}.

In the nascent stages of UAM system development, vertiport site selection is particularly important. Urban areas are expected to undergo significant changes with the introduction of UAM and vertiports, necessitating a comprehensive approach that takes into account for various urban factors such as operational, social and economic factors. The Community Integration Working Group (CIWG) highlights extensive considerations on the integration of UAM into urban environments, and NASA has organized these insights into a comprehensive list of 18 main themes of considerations for vertiport placement \citep{mendonca2022advanced}. These themes underscore the multifaceted nature of the challenge, highlighting the need for a comprehensive framework that addresses various considerations.

Existing studies on vertiport site selection have employed various methodologies such as clustering algorithms, simulation-based methods, custom optimization models, expert surveys, and GIS-based approaches. These approaches have provided valuable insights and laid a strong foundation for understanding this complex problem. Nevertheless, given the intricate dynamics of urban environments, there remains an opportunity to expand upon these efforts by incorporating more comprehensive perspectives on urban environments.

To contribute to this growing body of work, we propose a novel framework leveraging Latent Factor Analysis (LFA) on regional data to uncover latent factors that characterize the urban environment for vertiport integration. These latent factors are grouped into two pillars: suitability and attractiveness.

\begingroup
\begin{itemize}
    \item \textit{Suitability} refers to the operational feasibility of a location for a vertiport. Factors such as community acceptance and the potential for minority gentrification are critical considerations for suitability. A region may be deemed unsuitable if there is strong community opposition or if the region is overly affluence, posing social justice barriers.
    \item \textit{Attractiveness} pertains to the operational merits of the vertiport. A region is deemed attractive if it is expected to capture significant existing demand or has operational advantages. Factors such as potential connectivity to nearby public transit systems or high economic activity contribute to a location’s attractiveness for a vertiport.
\end{itemize}
\endgroup

In essence, suitability is a “must-have” - a basic requirement for vertiport placement, while attractiveness is a “nice-to-have” - factors that enhance the potential success and integration of the vertiport. Together, these pillars form the foundation for effective vertiport site selection in complex urban settings.

LFA emerges as a critical solution, uncovering and quantifying latent factors that reflect underlying urban structures and dynamics. This approach enables a comprehensive evaluation that integrates social considerations like community preparedness with economic factors such as economic dynamism. By consolidating these various factors into a composite v-score metric, LFA offers a holistic perspective on urban environments, ensuring sustainable vertiport integration that balances operational feasibility with operational merits. 

This paper is structured as follows: Section 2 provides details on existing work with diverse methodologies. Section 3 introduces the framework and methodology. Section 4 presents data and results for Seoul case study, and Section 5 discussion further result analysis. Finally, Section 6 concludes the paper with suggestions for future studies.

\section{Literature Review}
Numerous studies have explored the problem of vertiport site selection using various methodologies. Expanding on the comprehensive literature review by \citet{long2023demand}, we further identified various approaches in three groups: demand estimation, expert opinions, and geographic information system approach.

\subsection{Demand Estimation}
\textbf{Clustering algorithms}, particularly K-means clustering, are widely used to identify vertiport locations. \citet{tarafdar2021comparative} expanded on the work by \citet{rimjha2020demand} and used the K-means algorithm and center-of-mass approaches for UAM landing site selection. \citet{rajendran2019insights} also applied K-means clustering on the predicted mobility data from the NYC Taxi and Limousine Commission data for the strategic placement of the vertiports in New York City. \citet{bulusu2021traffic} developed a traffic demand analysis method to identify areas with a high potential for UAM demand and used iterative K-means clustering to select the vertiport sites in the San Francisco Bay Area. 

\textbf{Simulation-based methods} involve tools such as MATSim, aiming to provide insights into UAM demand prediction and identify areas with high potential demand as vertiport locations. Articles by \citet{rothfeld2018agent} and \citet{balac2019prospects} introduced an approach utilizing agent-based simulations to assess the demand for UAM and determine optimal vertiport sites with travel patterns, user behaviors, and infrastructure constraints. \citet{wu2021integrated} used the simulation output of the Tampa Bay Regional Planning Model (TBRPM) along with socio-demographic data to explore interactions between UAM travel demand and vertiport locations. Building upon this work, \citet{zhao2022environmental} examined the environmental impacts of UAM implementation in the Tampa Bay area to provide insights into sustainable vertiport planning and location optimization.

\textbf{Optimization models} are employed by many researchers, tailored to predict UAM demand and optimize vertiport locations. \citet{ploetner2020long} predicted UAM demand in Europe based on possible adoption rates of UAM, incorporating savings in travel time, willingness to pay, and operational constraints to optimize vertiport locations that maximize demand coverage. \citet{rath2022air} focused on NYC Taxi users, developing a p-hub median location problem that maximizes travel time savings to multiple airports while incorporating factors such as user choice behavior and demand elasticity. \citet{chae2023vertiport} identified vertiport locations based on geographic and urban constraints, analyzing population density, commuter patterns, and accessibility to maximize demand coverage and ensure equitable access across the region.

\subsection{Expert surveys and opinions}
The expert surveys and opinions are employed through methods like the Analytical Hierarchy Process (AHP) and qualitative evaluation to weight various criteria for vertiport site selection. \citet{fadhil2018gis}  utilized a two-step AHP-Delphi method, engaging 15 experts to evaluate and weight the socioeconomic factors influencing site selection. Similarly, \citet{lee2023proposal} utilized the AHP method in their studies. They integrated aviation expert knowledge into spatial data analysis, addressing airspace regulations, obstacle clearance requirements, and noise considerations for UAM route design and vertiport placement. \citet{preis2021quick} incorporated expert interviews and workshop discussions to formulate criteria for vertiport site selection and anticipate effects. \citet{nordstrom2022optimal} combined expert knowledge with public surveys and analytical approaches to identify optimal vertiport locations. Furthermore, \citet{cho2022can} explored UAM's potential in Seoul by analyzing socio-demographic factors and public adoption readiness. Their study integrated expert insights to address challenges such as safety, noise, and infrastructure requirements.

\subsection{Geographic information system (GIS) approach}
Geographical information system (GIS) approach is utilized to integrate multiple urban factors into spatial analysis for vertiport site selection. \citet{wu2021integrated} incorporated physical and regulatory factors for UAM operations through a three-dimensional GIS map from lidar data to identify candidate locations in Florida. \citet{sinha2022novel} expanded on \citet{rajendran2019insights}, employing a multi-criteria warm start technique that considers socioeconomic variables as layers and utilized the multi-criteria decision-making technique to compute the total score of suggested sites. \citet{sheth2023vamos} developed a regional modeling tool named VAMOS!, which utilizes numerical score inputs from users for vertiport placement, including various attributes like zoning, environmental impact, and intermodal transportation systems as layers. \citet{wei2023land} conducted a case study on vertiport placement in the San Francisco Bay Area, developing a systematic framework using GIS. Their study identified three key parameters\(-\)safety, access, and equity\(-\)to evaluate site suitability.  Additionally, \citet{yoon2025integrating} proposed a strategic GIS-based methodology to integrate UAM with highway infrastructure in the Seoul Metropolitan Area, evaluating 148 candidates based on geographic information, origin-destination volume, and travel time to identify 56 optimal vertiport locations.

Many studies primarily prioritized demand estimation and economic aspects, neglecting the complex nature of UAM integration in urban environments. Expert-driven analyses, while providing valuable insight, are particularly restrictive given the nascent stage of vertiport development and the evolving nature of UAM systems. Although GIS approaches are capable of accounting for different dimensions within the urban area, these studies often analyze data independently, overlooking the interconnected interplay of urban factors.

\section{Methodology}

\begin{figure}[!ht]
  \centering
  \includegraphics[width=1\textwidth]{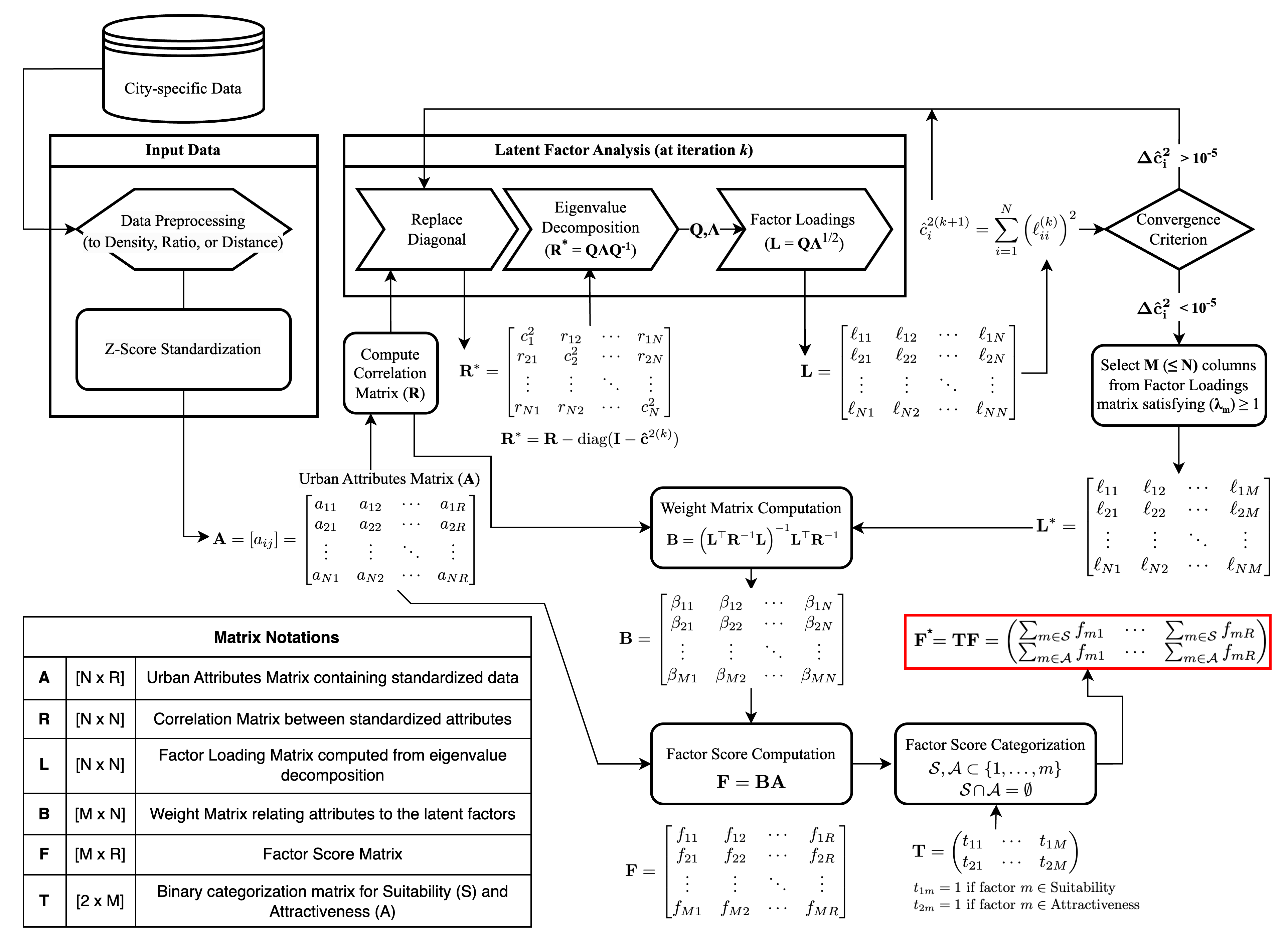}
  \caption{Flow of USE-LFA, from the city-specific urban attributes to the suitability and attractiveness scores for each region.}\label{fig:framework}
\end{figure}

We propose the Urban Site Evaluation using Latent Factor Analysis (USE-LFA) framework, a data-driven approach that captures the multifaceted nature of urban environments by disentangling them into two distinct dimensions: suitability and attractiveness. USE-LFA applies latent factor analysis (LFA) to regional urban data to uncover latent structures that characterize urban environments relevant to vertiport integration (see Figure \ref{fig:framework}). This study employs principal axis factoring (PAF) with Varimax rotation and regression-based scoring to identify latent factors underlying urban attributes across regions. 

\subsection{Input Data Preparation for USE-LFA}
The table below provides an example of data attributes that can be used as input for USE-LFA. The vertiport-related themes are predetermined and relevant data is gathered accordingly. The input attributes are flexible and can be tailored to suit the unique characteristics of each study area. The description column outlines the normalization process, an essential step for LFA to address variations in census tract sizes and population densities across the regions.

\begin{table}[!ht]
    \caption{Input Description of the Urban Attributes}
    \label{tab:data_corpus}
    \begin{center}
        \resizebox{\textwidth}{!}{%
            \begin{tabular}{>{\raggedright\arraybackslash}p{3cm} >{\raggedright\arraybackslash}p{1cm} >{\raggedright\arraybackslash}p{5cm} >{\raggedright\arraybackslash}p{9cm}}
                \toprule
                \textbf{Theme} & \textbf{N} & \textbf{Urban Attributes} & \textbf{Data Description} \\ 
                \midrule
                \textbf{Population} 
                & 1  & Housing Density                 & Count of housing units per km$^2$ (units/km$^2$). \\ 
                & 2  & Floating Population             & Hourly floating population (8 AM–8 PM) density (pax/km$^2$). \\ 
                & 3  & Foreign Population              & Hourly foreign floating population density (pax/km$^2$). \\ 
                & 4  & Commuting Population            & Ratio of morning commute flows to the registered population. \\ 
                & 5  & Incoming Population             & Ratio of inbound to outbound commuters during peak hours. \\ 
                & 6  & Disabled Population             & Percentage of disabled residents relative to the total population (\%). \\ 
                & 7  & Beneficiary                     & Welfare recipients as a percentage of the census tract population (\%). \\ 
                & 8  & Senior Population               & Proportion of elderly residents (65+) in the registered population (\%). \\ 
                \cmidrule(lr){1-4}
                \textbf{Socioeconomic} 
                & 9  & Average Income                  & Monthly average income aggregated at the census level (KRW/month). \\ 
                & 10 & Spending Ratio                  & Total expenditures as a percentage of monthly average income (\%). \\ 
                & 11 & Transportation Expenditure & Transit costs (bus, express bus, taxi, train, and metro) as a percentage of monthly income (\%). \\ 
                \cmidrule(lr){1-4}
                \textbf{Transport Infrastructure} 
                & 12 & Bus Stop Density                & Total bus stops per km$^2$, calculated via spatial joins with administrative boundaries (stops/km$^2$). \\ 
                & 13 & Metro Coverage                  & Percentage of land within a 1,000m buffer around metro stations (\%). \\ 
                & 14 & Metro Users Per Station     & Daily metro users per station in a census tract during weekdays (users/station). \\ 
                & 15 & Bus Users Per Stop       & Daily bus boardings per bus stop in a census tract during weekdays (users/stop). \\ 
                \cmidrule(lr){1-4}
                \textbf{Land Use and Facilities}
                & 16 & Parking Space Density           & Parking spaces per km$^2$ based on POI data (spaces/km$^2$). \\ 
                & 17 & Touristic Facility Density      & Count of touristic facilities per km$^2$ from POI data (facilities/km$^2$). \\ 
                & 18 & Cultural Facility Density       & Cultural facilities per km$^2$, spatially aggregated (facilities/km$^2$). \\ 
                & 19 & Energy Usage Intensity          & Annual energy consumption normalized by area (kWh/km$^2$/year). \\ 
                & 20 & Employment Density              & Total employees per km$^2$ based on business registries (employees/km$^2$). \\ 
                & 21 & Land Price                      & Land value per km$^2$ from official surveys (KRW/km$^2$). \\ 
                \cmidrule(lr){1-4}
                \textbf{Spatial Proximity} 
                & 22 & Distance to Helipads           & Euclidean distance from census centroid to the nearest helipad location (km). \\ 
                & 23 & Distance to Hospitals          & Euclidean distance from census centroid to the nearest hospital location (km). \\ 
                & 24 & Distance to VFR Routes         & Euclidean distance from census centroid to the nearest VFR route (km). \\ 
                & 25 & Distance to Fire Stations        & Euclidean distance from census centroid to the nearest fire station (km). \\ 
                \bottomrule
            \end{tabular}%
        }
    \end{center}
\end{table}

\subsection{Z-Score Standardization}
Due to the framework's sensitivity to data scale, Z-score standardization was applied to the input data. This process ensures that attributes with larger variances do not overshadow others, enabling a more balanced and accurate analysis.

\subsection{Latent Factor Analysis (LFA)}
The core constitution of USE-LFA is a robust mathematical model that decomposes complex regional data into interpretable factors. Let $\mathbf{A} \in \mathbb{R}^{N \times R}$ represent the standardized urban attribute matrix where $N$ is the number of standardized urban attributes (indexed by $i=1,2,\dots,N$) and $R$ is the number of regions in the study area (indexed by $j=1,2,\dots,R$). Each element $a_{ij}$ encodes the standardized measurement of urban attribute $i$ in region $j$. The foundational relationship between the urban attributes and their underlying latent factors is formalized as:
\begin{equation}
    \boldsymbol{A} = \mathbf{L}\mathbf{F} + \mathbf{E}.
\end{equation}
Where $\mathbf{L} \in \mathbb{R}^{N \times M}$ represents the urban attribute loading matrix with elements $\ell_{im}$ capturing the intensity of relationships between urban attribute $i$ and latent factor $m$. The factor score matrix $\mathbf{F} \in \mathbb{R}^{M \times R}$ contains elements $f_{mj}$ indicating the degree to which region $j$ embodies latent urban factor $m$. The matrix product $\mathbf{LF}$ can be decomposed as $\sum_{m=1}^{M} \boldsymbol{\ell}_m \mathbf{f}_m$, where $\boldsymbol{\ell}_m = [\ell_{1m}, \ldots, \ell_{Nm}]^{\top} \in \mathbb{R}^N$ represents the loading vector for factor $m$ across all attributes, and $\mathbf{f}_m = [f_{m1}, \ldots, f_{mR}] \in \mathbb{R}^R$ represents the score vector for factor $m$ across all regions. Finally, $\mathbf{E} \in \mathbb{R}^{N \times R}$ captures region-specific variance unique to each attribute. This matrix factorization approach enables dimensionality reduction while preserving the essential structure of urban environments, allowing meaningful latent factors to be discovered from regional input urban attributes.

\subsection*{Initial Communality Estimation}
Principal Axis Factoring (PAF) is used to extract the urban latent factors by isolating the shared variance among the urban attributes. The process undergoes an iterative procedure that refines estimates of communalities$-$the portion of each attribute's variance explained by the latent factors.

For standardized input urban attribute matrix $\mathbf{A}$, the covariance structure reduces to:
\begin{equation}
    \text{Cov}(\boldsymbol{A}_{standardized}) = \text{Corr}(\boldsymbol{A}_{raw}) =\mathbf{R},
\end{equation}
where $\mathbf{R}$ hereby indicates the correlation matrix of the original (non-standardized) urban attributes. This equivalence enables the direct interpretation of factor loadings as standardized coefficients and scale-invariant comparison of urban attribute relationships. On correlation matrix $\mathbf{R}$, the initial communality $\hat{\mathbf{c}}_i^{2(0)}$ for urban attribute $i$ is estimated via squared multiple correlations:
\begin{equation}
    \hat{\mathbf{c}}_i^{2(0)}=1-{\frac{1}{r^{ii}}},
\end{equation}
where $r^{ii}$ is the $i^{th}$ diagonal element of the inverse correlation matrix $\mathbf{R^{-1}}$. This initialization step approximates how much variance in urban attribute $i$ is explained by its linear relationship with other attributes. 

\subsection*{Iterative Procedure}

\paragraph{Adjusted Correlation Matrix}
The iterative factor extraction process subsequently operates on an \textit{adjusted} correlation matrix $\mathbf{R}^*$, where diagonal elements (original correlations of 1) are replaced with communality estimates to focus on shared variance among the urban attributes:
\begin{equation}
    \forall i: \mathbf{R^*}=\mathbf{R}-\text{diag}(\mathbf{I}_i-\mathbf{\hat{c}}_i^{2(0)}).
\end{equation}

\paragraph{Eigenvalue Decomposition}
The adjusted correlation matrix $\mathbf{R}^*$ is decomposed via eigenvalue decomposition:
\begin{equation}
    \mathbf{R}^* = \mathbf{Q}\mathbf{\Lambda}\mathbf{Q}^\top \quad \text{with} \quad 
    \begin{cases}
    \mathbf{Q} = [\mathbf{q}_1|\cdots|\mathbf{q}_N] \in \mathbb{R}^{N \times N} \\
    \mathbf{\Lambda} = \text{diag}(\lambda_1, \ldots, \lambda_N)
    \end{cases}
\end{equation}
where $\mathbf{Q}$ represents the eigenvectors with $\mathbf{q}_m = [q_{1m}, ..., q_{Nm}]^\top \in \mathbb{R}^N$ and $\mathbf{\Lambda}$ contains the corresponding eigenvalues.

\paragraph{Factor Loading Estimation}
At each iteration $k$, the urban factor loadings are computed as follows:
\begin{equation}
    \mathbf{L}^{(k)} = \mathbf{Q}_N^{(k)} \left(\mathbf{\Lambda}_N^{(k)}\right)^{1/2} \quad \in \mathbb{R}^{N \times N}
\end{equation}
where $\mathbf{Q}_N^{(k)}$ contains the eigenvectors and $\mathbf{\Lambda}_N^{(k)}$ the corresponding eigenvalues at iteration $k$.

\paragraph{Updating Communalities}
Using the current loading matrix $\mathbf{L}^{(k)}$, updated communalities are calculated as:
\begin{equation}
    \hat{c}_i^{2(k+1)} = \sum_{m=1}^M \left(\ell_{im}^{(k)}\right)^2 
\end{equation}
where $\ell_{im}^{(k)}$ denotes the loading of the urban attribute $i$ on factor $m$ at iteration $k$. These values represent the cumulative explained variance for each attribute across all latent urban factors. The updated communalities $\hat{c}_i^{2(k+1)}$ then replace the diagonal entries of the correlation matrix, generating the adjusted matrix $\mathbf{R}^{*(k+1)}$ for the subsequent iteration $k+1$.

\paragraph{Convergence Criterion}
The iteration terminates when the total communality change reaches the threshold:
\begin{equation}
\sum_{i=1}^N \left|\hat{c}_i^{2(k+1)} - \hat{c}_i^{2(k)}\right| < \epsilon \quad (\epsilon = 10^{-5})
\end{equation}
Final factor loadings $\mathbf{L}^*$ are obtained at the convergence iteration $k_{\text{final}}$. Following the Kaiser criterion, only factors with eigenvalues $\lambda_m \geq 1$ are retained, ensuring each factor explains at least as much variance as a single original attribute.

The urban factor loadings in $\mathbf{L}^*$ can be interpreted as correlation coefficients between each urban attribute and the extracted factors, with larger absolute values indicating stronger relationships. These loadings are essential for identifying which urban characteristics cluster together to form meaningful latent dimensions of urban environment.

\subsection*{Factor Rotation}
To enhance interpretability, the Varimax rotation is implemented on the factor structure. The Varimax rotation maximizes the sum of variances of squared loadings within each factor, which tends to produce a simpler, more interpretable factor structure with loadings closer to either 0 or 1. The rotated factor loading matrix is given by:
\begin{equation}
    \mathbf{L}^R =  \mathbf{L}^{*}\mathbf{V},
\end{equation}
where $\mathbf{V}\in \mathbb{R}^{M\times M}$ denotes the orthogonal Varimax rotation matrix satisfying $\mathbf{V}^\top\mathbf{V} = \mathbf{I}_M$. 

\subsection*{Factor Score Estimation}
Regional urban factor scores were computed via the regression estimator:
\begin{equation}
    \mathbf{B} = \Bigl( \mathbf{L}^\top \mathbf{R}^{-1} \mathbf{L}\Bigr)^{-1} \mathbf{L}^\top \mathbf{R}^{-1}
    \quad \in \mathbb{R}^{M \times N},
\end{equation}
where $\mathbf{R} \in \mathbb{R}^{N \times N}$ is the original correlation matrix and $\mathbf{L}$ refers to the rotated factor loading matrix $\mathbf{L}^R$. The factor score matrix $\mathbf{F} \in \mathbb{R}^{M \times R}$ is then obtained through:
\begin{equation}
\mathbf{F} = \mathbf{B}\mathbf{A} = \left[\sum_{i=1}^N \beta_{mi}a_{ij}\right],
\end{equation}
with component-wise form:
\begin{equation}
f_{mj} = \sum_{i=1}^N \beta_{mi}a_{ij} \quad \forall m \in {1,\ldots,M}, \ j \in {1,\ldots,R}.
\end{equation}
$a_{ij} \in \mathbb{R}$ represents the standardized value of urban attribute $i$ in region $j$. The summation aggregates the contributions from all $N$ urban attributes in region $j$, each weighted by the corresponding $\beta$ coefficient, to yield latent factor scores for a region.

The factor score is then divided into suitability and attractiveness, through the following composite definition matrix $\mathbf{T} \in \{0,1\}$:
\begin{equation}
    \mathbf{T} = \begin{pmatrix}
    t_{11} & \cdots & t_{1M} \\
    t_{21} & \cdots & t_{2M}
\end{pmatrix}
\end{equation}
where $t_{1m}=1$ if factor $m \in \text{Suitability}$, $t_{2m}=1$ if factor $m \in \text{Attractiveness}$, with all other entries being $0$.

The composite score matrix $\hat{\mathbf{F}}$ is computed through:
\begin{equation}
    \hat{\mathbf{F}} = \mathbf{T}\mathbf{F} = \begin{pmatrix}
    \sum_{m \in \mathcal{S}} f_{m1} & \cdots & \sum_{m \in \mathcal{S}} f_{mR} \\
    \sum_{m \in \mathcal{A}} f_{m1} & \cdots & \sum_{m \in \mathcal{A}} f_{mR}
    \end{pmatrix}
    \quad \in \mathbb{R}^{2 \times R},
\end{equation}
where $\mathcal{S}$ = Suitability factor indices and $\mathcal{A}$ = Attractiveness factor indices, such that $\mathcal{S,A} \subset \{1, \dots, m\}$ and $\mathcal{S} \cap \mathcal{A} = \emptyset$.

The framework distills complex urban attributes into interpretable dimensions by employing latent factor analysis to identify critical urban factors, and then strategically categorizing these factors into Suitability and Attractiveness. This binary categorization offers a structured analytics of a dual-perspective evaluation of urban regions while acknowledging the multidimensional nature of urban environment. The framework's mathematical formalization ensures reproducibility across different urban contexts, while its flexibility allows researchers and planners to adapt factor classifications based on specific policy objectives or theoretical considerations.

\section{Case study in Seoul}
The framework was demonstrated for Seoul as a case study. Globally recognized as one of the most UAM-ready regions \citep{KPMG_ATRI_2023}, Seoul has an extensive helipad infrastructure and has been the subject of numerous studies examining its potential for UAM implementation \citep{cho2022can, hwang2023study, yoon2025integrating}.

\subsection{Data Description}
Our analysis incorporated diverse data from Seoul to ensure a comprehensive evaluation. NASA has reportedly suggested 18 groups of themes with more than 450 considerations for vertiport placement \citep{mendonca2022advanced}. From these, 25 relevant regional urban attributes were selected for the vertiport site selection in Seoul (see Table \ref{tab:case_study_data}). The selected data was obtained from the Seoul Big Data Campus and the Seoul Open Data Plaza \citep{ SMG_BigDataCampus, SMG_SeoulOpenData}. These platforms provide open-source data that is accessible to anyone through their respective websites.

\begin{table}[!ht]
    \caption{REVISE: Data Attributes and Preprocessing for Case Study of Seoul}
    \label{tab:case_study_data}
    \centering
    % Use a smaller font and reduced spacing for a more compact table
    \small
    \setlength{\tabcolsep}{3pt}       % reduce horizontal gap between columns
    \renewcommand{\arraystretch}{0.8}  % reduce vertical spacing between rows
    \resizebox{\textwidth}{!}{%
        \begin{tabular}{%
          >{\raggedright\arraybackslash}p{0.5cm} 
          >{\raggedright\arraybackslash}p{4cm} 
          >{\raggedright\arraybackslash}p{3.5cm} 
          >{\raggedright\arraybackslash}p{4cm} 
          >{\raggedright\arraybackslash}p{5cm}}
            \toprule
            \textbf{N} & \textbf{Urban Attribute} & \textbf{Abbreviation} & \textbf{Data Source} & \textbf{Acquisition Process} \\ 
            \midrule
            1  & Housing Density           & HOUS\_DEN       & Seoul Housing Statistics by Neighborhood            & Areal weighting interpolation using census boundaries \\ 
            2  & Bus Stop Density          & BUS\_ST\_DEN    & Smart Card Transit Location Database                   & Spatial density calculation with 500m grid aggregation \\ 
            3  & Floating Population       & FL\_POP\_DEN    & Seoul Hourly Population Statistics                   & Mobile signaling data aggregation (8AM-8PM average) \\ 
            4  & Foreign Population        & FOR\_POP\_DEN   & Seoul Short-term Resident Population Data             & Immigration record filtering and temporal alignment \\ 
            5  & Commuting Population      & CMM\_POP\_RATIO & Seoul Mobility Pattern Survey                          & Morning commute ratio calculation (7-9AM) \\ 
            6  & Incoming Population       & I\_POP\_RATIO   & Seoul Mobility Pattern Survey                          & Peak hour inflow-outflow ratio computation \\ 
            7  & Parking Space Density     & PS\_DEN         & Kakao Map POI Database                                 & POI classification and kernel density estimation \\ 
            8  & Metro Coverage            & METRO\_COV\_P   & Seoul Metro Station Master Data                      & 1,000m buffer analysis around station centroids \\ 
            9  & Disabled Population       & DIS\_POP\_P     & Seoul Disability Registry                              & Population proportion calculation per administrative district \\ 
            10 & Beneficiary               & BEN\_POP\_P     & Seoul Basic Livelihood Security Registry               & Welfare recipient percentage computation \\ 
            11 & Senior Population         & SEN\_POP\_P     & Seoul Elderly Population Statistics                    & Age-based demographic proportion analysis \\ 
            12 & Average Income            & AVG\_INC        & Seoul Commercial District Analysis System              & Neighborhood-level income aggregation \\ 
            13 & Touristic Facility Density& TOUR\_DEN       & Kakao Map POI Database                                 & Tourism facility identification and density mapping \\ 
            14 & Cultural Facility Density & CULT\_DEN       & Kakao Map POI Database                                 & Cultural facility identification and density mapping \\ 
            15 & Energy Usage Intensity    & EN\_USE\_INT    & Seoul Energy Consumption Report                        & Annual kWh normalization by land area \\ 
            16 & Employment Density        & EMPL\_DEN       & Seoul Business Establishment Registry                  & Workforce density calculation per km² \\ 
            17 & Bus Users Per Stop        & BUS\_USR\_DEN   & Seoul Public Transit Ridership Data                   & Boarding counts normalized by stop locations \\ 
            18 & Metro Users Per Station   & METRO\_USR\_DEN & Seoul Metro Ridership Statistics                       & Station-level usage intensity calculation \\ 
            19 & Spending Ratio            & SPND\_P        & Seoul Commercial District Analysis System              & Expenditure-to-income ratio computation \\ 
            20 & Land Price                & LND\_PR        & Seoul Official Land Price Index                        & Parcel value aggregation per administrative district \\ 
            21 & Transportation Expenditure& TR\_EXP\_P     & Seoul Commercial District Analysis System              & Transit cost percentage calculation from income data \\ 
            22 & Distance to Helipads      & HELI\_DIST     & National Aviation Facility Registry                    & Euclidean distance from census tract centroids \\ 
            23 & Distance to Hospitals     & HOSP\_DIST     & National Medical Institution Database                  & Straight-line distance calculation \\ 
            24 & Distance to VFR Routes    & VFR\_DIST      & Aeronautical Navigation Chart System                   & Air corridor buffer analysis \\ 
            25 & Distance to Fire Stations & FS\_DIST       & Emergency Service Location Database                    & Service area proximity modeling \\ 
            \bottomrule
        \end{tabular}%
    }
    \renewcommand{\arraystretch}{1} % Reset row spacing if needed
    \hspace*{\fill}{\scriptsize Note: All metrics at census tract level. VFR = Visual Flight Rules. GIS processing applied.}
\end{table}

The descriptive statistics table (Table \ref{tab:desc_stats}) summarize 426 observations across 25 urban attributes selected for vertiport site selection in Seoul. The data reveals substantial variability in urban characteristics, with several metrics showing notable statistical properties. Population-related measures (e.g., FL\_POP\_DEN, FOR\_POP\_DEN) and specialized facilities (e.g., CULT\_DEN) exhibit right-skewed distributions, indicating concentrated density in specific areas. Infrastructure coverage metrics like METRO\_COV\_P demonstrate more uniform distribution across the city. These distribution patterns highlight the heterogeneity of Seoul's urban landscape, with implications for optimal vertiport placement. The data's non-normal distribution characteristics informed our subsequent analytical approach using principal axis factoring rather than standard methods requiring normality assumptions.
\begin{table}[!ht]
    \caption{Descriptive Statistics of Raw Urban Attributes}
    \label{tab:desc_stats}
    \centering
    \resizebox{\textwidth}{!}{%
        \begin{tabular}{lrrrrrrrr}
            \toprule
            \textbf{Variable} & \textbf{Count} & \textbf{Mean} & \textbf{Std} & \textbf{Min} & \textbf{Median} & \textbf{Max} & \textbf{Skewness} & \textbf{Kurtosis} \\
            \midrule
            HOUS\_DEN          & 426 & 7109.123   & 3811.969   & 0.000   & 6824.086   & 20192.857  & 0.407  & -0.001 \\
            FL\_POP\_DEN       & 426 & 23216.819  & 12506.621  & 0.000   & 22167.694  & 100436.473 & 1.011  & 3.210  \\
            FOR\_POP\_DEN      & 426 & 347.584    & 1061.019   & 0.000   & 83.023     & 13992.351  & 8.002  & 81.075 \\
            CMM\_POP\_RATIO    & 426 & 4445.137   & 2485.035   & 0.000   & 4058.770   & 15616.181  & 0.784  & 0.960  \\
            I\_POP\_RATIO      & 426 & 0.970      & 0.833      & 0.000   & 0.744      & 7.033      & 3.741  & 18.591 \\
            DIS\_POP\_P        & 426 & 0.041      & 0.017      & 0.000   & 0.040      & 0.174      & 2.498  & 14.291 \\
            BEN\_POP\_P        & 426 & 0.040      & 0.031      & 0.000   & 0.036      & 0.236      & 2.286  & 9.521  \\
            SEN\_POP\_P        & 426 & 0.182      & 0.040      & 0.000   & 0.181      & 0.359      & 0.195  & 1.837  \\
            AVG\_INC           & 426 & 3.308      & 1.099      & 0.000   & 3.101      & 7.421      & 0.249  & 2.164  \\
            SPEND\_P          & 426 & 2128.242   & 10895.023  & 0.000   & 477.260    & 140851.873 & 9.826  & 105.405 \\
            TR\_EXP\_P         & 426 & 0.067      & 0.107      & 0.000   & 0.026      & 0.908      & 3.449  & 16.445 \\
            BUS\_ST\_DEN       & 426 & 24.289     & 12.967     & 0.000   & 22.518     & 84.783     & 1.003  & 1.639  \\
            METRO\_COV\_P      & 426 & 0.662      & 0.319      & 0.000   & 0.754      & 1.000      & -0.651 & -0.862 \\
            METRO\_USR\_DEN    & 426 & 5538.780   & 11683.466  & 0.000   & 7351.026   & 110019.000 & 3.640  & 20.021 \\
            BUS\_USR\_DEN      & 426 & 465.426    & 268.217    & 0.000   & 386.614    & 1397.289   & 1.064  & 0.686  \\
            PS\_DEN            & 426 & 9794.966   & 5851.015   & 0.000   & 9363.484   & 70584.091  & 3.041  & 26.676 \\
            TOUR\_DEN          & 426 & 1.620      & 2.489      & 0.000   & 1.102      & 28.085     & 5.045  & 39.433 \\
            CULT\_DEN          & 426 & 4.636      & 13.141     & 0.000   & 1.588      & 206.410    & 9.905  & 135.248 \\
            EN\_USE\_INT       & 426 & 1.000      & 1.001      & 0.032   & 0.688      & 8.540      & 3.826  & 20.183 \\
            EMPL\_DEN          & 426 & 11269.860  & 12975.330  & 0.000   & 7248.035   & 109586.870 & 3.378  & 15.490 \\
            LAND\_PR           & 426 & 4.663      & 3.092      & 0.000   & 3.802      & 23.770     & 2.284  & 6.917  \\
            HELI\_DIST         & 426 & 1.946      & 1.486      & 0.013   & 1.534      & 8.109      & 1.187  & 1.274  \\
            HOSP\_DIST         & 426 & 3.421      & 2.073      & 0.040   & 2.931      & 10.524     & 0.987  & 0.727  \\
            VFR\_DIST          & 426 & 3.103      & 2.169      & 0.005   & 2.808      & 8.751      & 0.537  & -0.667 \\
            FS\_DIST           & 426 & 2.037      & 1.143      & 0.000   & 1.846      & 7.244      & 0.959  & 1.949  \\
            \bottomrule
        \end{tabular}%
    }
\end{table}

\subsection{LFA Results}
Based on Kaiser's criterion, six factors were retained from the 25 urban attributes, collectively accounting for 66.82\% of the total cumulative variance. 
\begin{table}[!ht]
    \caption{Resultant Factors, Eigenvalues, and Variances from LFA}
    \label{tab:initial_result}
    \begin{center}
        \begin{tabular}{rrrr}
            \toprule
            \textbf{Factor} & \textbf{Initial Eigenvalues} & \textbf{\% of Variance} & \textbf{Cumulative Variance} \\
            \midrule
            1 & 3.405219 & 13.620877 & 13.620877 \\
            2 & 3.847866 & 15.391465 & 29.012342 \\
            3 & 2.946635 & 11.786539 & 40.798881 \\
            4 & 2.506135 & 10.024542 & 50.823423 \\
            5 & 2.742083 & 10.968330 & 61.791753 \\
            6 & 1.256834 & 5.027335 & 66.819088 \\
            \bottomrule
        \end{tabular}
    \end{center}
\end{table}
Table \ref{tab:initial_result} details the eigenvalue of each factor, representing the variance in the original variables explained by that factor. The first factor had an eigenvalue of 3.41, accounting for 13.62\% of the total variance. Factor 2 had the highest eigenvalue of 3.85, which explained 15.39\% of the total variance. In total, the six identified factors explain 66.82\% of the total variance. These results indicate that a relatively small set of underlying factors can account for a substantial portion of the variance in the original attributes, supporting their use for further analysis and interpretation.

\begin{figure}
    \centering
    \includegraphics[width=0.9 \linewidth]{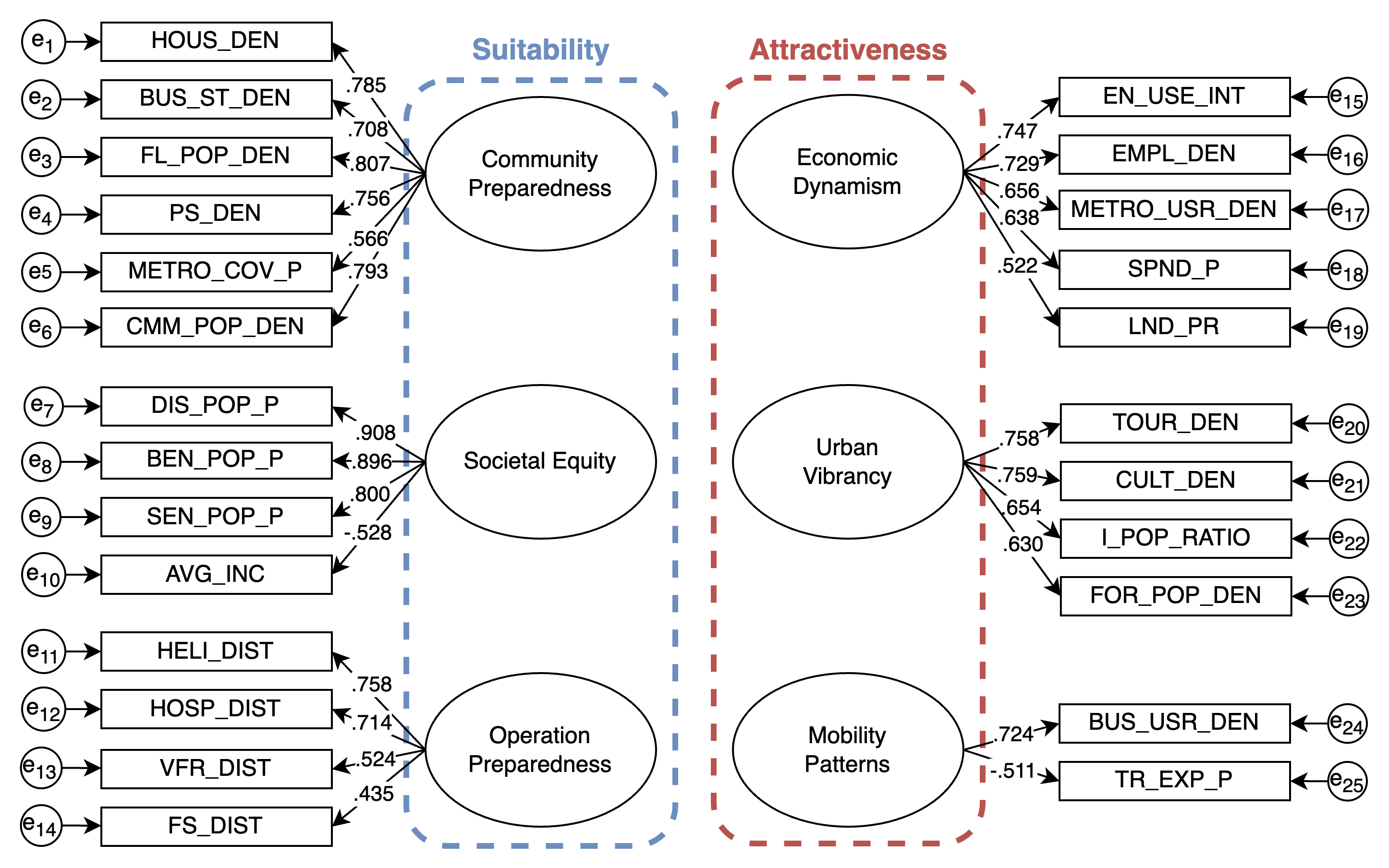}
    \caption{Structural LFA results showing the dominant attributes for each latent factor.}\label{fig:LFA_results}
\end{figure}

For each attribute, the factor with the highest absolute factor loading was identified, designating the attribute as a “dominant attribute” for that factor. These dominant attributes, grouped within the same factor, highlight shared characteristics that define the factor (see Figure \ref{fig:LFA_results}). Based on the overarching characteristics, each factor was interpreted (see Table \ref{tab:factor_result}). The impact directions were inferred on the basis of the associated variables and their anticipated influences on the vertiport placement.
\begin{table}[!ht]
    \caption{Factor interpretations and factor loadings}
    \label{tab:factor_result}
	\begin{center}
        \resizebox{\textwidth}{!}{
            \begin{tabular}{|c|c|c|c|c|c|c|}
                \hline
                \textbf{Latent Factors} & \textbf{Impact} & \textbf{Dominant Attributes} & \textbf{Factor Loadings} & \textbf{Eigenvalue} & \textbf{\% Of Variance} \\
                \hline

                1. Economic Dynamism & (+) 
                    & EN\_USE\_INT    & 0.747 & 3.41 & 13.62 \\
                 &  & EMPL\_DEN       & 0.729 &  &  \\
                 &  & METRO\_USR\_DEN & 0.656 &  &  \\
                 &  & SPND\_P         & 0.638 &  &  \\
                 &  & LND\_PR         & 0.522 &  &  \\
                \hline
                
                2. Community Preparedness & (-) 
                    & HOUS\_DEN     & 0.785 & 3.85 & 15.39 \\
                 &  & BUS\_ST\_DEN  & 0.708 &  &  \\
                 &  & FL\_POP\_DEN  & 0.807 &  &  \\
                 &  & PS\_DEN       & 0.756 &  &  \\
                 &  & METRO\_COV\_P & 0.793 &  &  \\
                \hline

                3. Societal Equity& (+) 
                    & DIS\_POP\_P & 0.908 & 2.95 & 11.79 \\
                 &  & BEN\_POP\_P & 0.896 &  &  \\
                 &  & SEN\_POP\_P & 0.800 &  &  \\
                 &  & AVG\_INC    & -0.528 &  &  \\
                \hline

                4. Operation Preparedness & (-) 
                    & HELI\_DIST & 0.758 & 2.51 & 10.02 \\
                 &  & HOSP\_DIST & 0.714 &  &  \\
                 &  & VFR\_DIST  & 0.524 &  &  \\
                 &  & FS\_DIST   & 0.435 &  &  \\
                \hline

                5. Urban Vibrancy & (+) 
                    & TOUR\_DEN     & 0.758 & 2.74 & 10.97 \\
                 &  & CULT\_DEN     & 0.759 &  &  \\
                 &  & I\_POP\_RATIO & 0.654 &  &  \\
                 &  & FOR\_POP\_DEN & 0.630 &  &  \\
                \hline

                6. Mobility Patterns & (+) 
                    & BUS\_USR\_DEN & 0.724 & 1.26 & 5.03 \\
                 &  & TR\_EXP\_P    & -0.511 &  &  \\
                \hline
            \end{tabular}
        }
    \end{center}
\end{table}

\subsection{Identified Factors into Suitability and Attractiveness}
Based on the LFA results, the framework establishes two essential composite measures: suitability (derived from factors 2, 3, and 4) and attractiveness (from factors 1, 5, and 6). These complementary metrics capture different aspects of vertiport viability - suitability addresses operational feasibility considerations, while attractiveness reflects operational merit indicators.

\subsection*{Suitability Factors}

\textbf{Factor 2: Community Preparedness} encompasses the density of housing units, bus stations, floating population, car parking spaces, and metro service coverage. This factor receives a negative weight to account for potential community resistance and challenges in the initial adoption of UAM. 

Concerns about ground population safety near vertiports have been noted by \citet{yedavalli2019assessment}, highlighting potential resistance from communities. \citet{oh2022data} stressed the importance of accounting for spatio-temporal variations in population density to assess and mitigate risks associated with UAS operations in urban areas. Additionally, \citet{smith2024supporting} emphasized the importance of public engagement and transparent communication about UAM's impacts to foster acceptance. Similar to the adverse effects of air transportation noise \citep{friedt2021perception}, frequent UAM operations could generate significant noise pollution, making densely populated areas less suitable for vertiport placement. \citet{oh2024urban} highlighted the need to balance operational feasibility with minimizing risks in densely populated regions through a Pareto-based framework for urban airspace assessment.

While established transit infrastructure may offer connectivity advantages \citep{straubinger2020overview, wang2023review, yan2024urban}, competition with existing transportation modes \citep{hasan2019urban}, induced traffic \citep{yadav2025modelling}, and safety risks \citep{al2020factors} could pose significant concerns during early UAM adoption\(-\)particularly in cities like Seoul. These considerations suggest that areas with high density and extensive transit connectivity may be less suitable for vertiport placement.

\textbf{Factor 3: Societal Equity} represents the demographic and socioeconomic attributes of metropolitan areas, including the ratios of disabled populations, national basic livelihood recipients, senior populations, and average income levels. These indicators help identify regions with concentrations of vulnerable populations, offering essential insights for suitability assessments that prioritize social justice.

\textbf{Factor 4: Operation Preparedness} is formulated based on proximity to critical emergency infrastructures such as fire stations, major hospitals, helipads, and VFR routes. These variables reflect a region's preparedness to meet operational requirements and support emergency responses. In contingency scenarios at vertiports, proximity to these critical facilities ensures faster reaction times, enhancing safety—particularly important for mitigating risks associated with battery-operated UAM aircraft \citep{wisk2022concept}. 

Additionally, close access to emergency services improves response capabilities in traffic-congested areas, where UAM vehicles can serve as vital reinforcements to the emergency response network. Existing helipads and VFR routes provide foundational infrastructure that can be leveraged during the initial stage of UAM implementation, further supporting operational suitability \citep{FAA_UTM_ConOps_2020}.

\subsection*{Attractiveness Factors}

\textbf{Factor 1: Economic Dynamism} is gauged through energy consumption relative to the average, working population density, weekday hourly metro users per station, spending-to-income ratio, and official land prices. These metrics collectively reflect the economic vibrancy of an area, establishing a foundation for evaluating attractiveness. This aspect is particularly important given the relatively high expected cost of UAM flights (\$3.50 to \$4.00 per passenger-km) during the initial stages of implementation \citep{binder2018if, rimjha2021urban}.

\textbf{Factor 5: Urban Vibrancy} evaluates the centrality of regions within Seoul, measured by the density of touristic and cultural facilities, the presence of foreigners, and the ratios of moving and incoming populations. These variables together indicate the attractiveness of regions by revealing their centrality and dynamism.

\textbf{Factor 6: Mobility Patterns} are assessed through weekday hourly bus users per stop and the inversely related ratio of transportation expenditures to total spending. Areas with high bus usage combined with low transportation spending suggest either higher reliance on personal vehicles or higher incomes that make public transportation expenses relatively insignificant. These regions with established travel demand and lower relative transportation costs indicate potential attractiveness for adopting new transportation modes like UAM.

\subsection{Suitability and Attractiveness Scores}

\begin{figure}[!ht]
  \centering
  \includegraphics[width=0.75\textwidth]{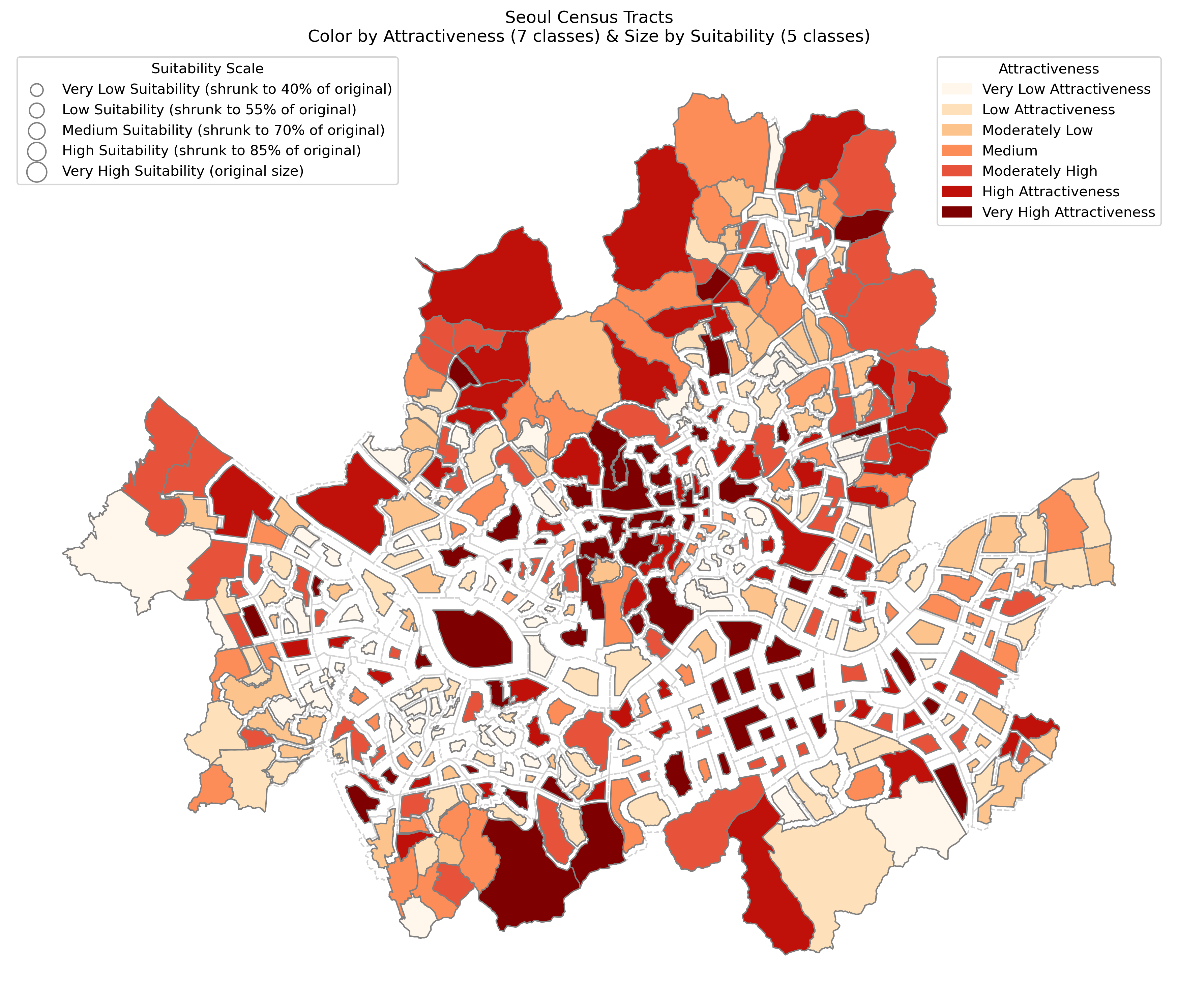}
  \caption{Spatial Distribution of Suitability and Attractiveness Scores in Seoul}\label{fig:SA_vismap}
\end{figure}

The framework categorizes urban factors into two dimensions critical for UAM site selection: suitability (operational feasibility) and attractiveness (operational merits). Figure \ref{fig:SA_vismap} illustrates their spatial distribution across Seoul, revealing distinct geographic patterns. Central business districts exhibit concentrated attractiveness scores, reflecting their inherent advantages in demand generation and economic activity. In contrast, suitability scores display broader dispersion, with elevated values observed in less congested peripheral regions. This dichotomy establishes a foundational tension between immediate operational viability and long-term scalability in vertiport planning.

\section{Result Analysis and Discussions}
\subsection{Comparative Analysis of Top-Scoring Regions}
\begin{figure}[!ht]
  \centering
  \includegraphics[width=1\textwidth]{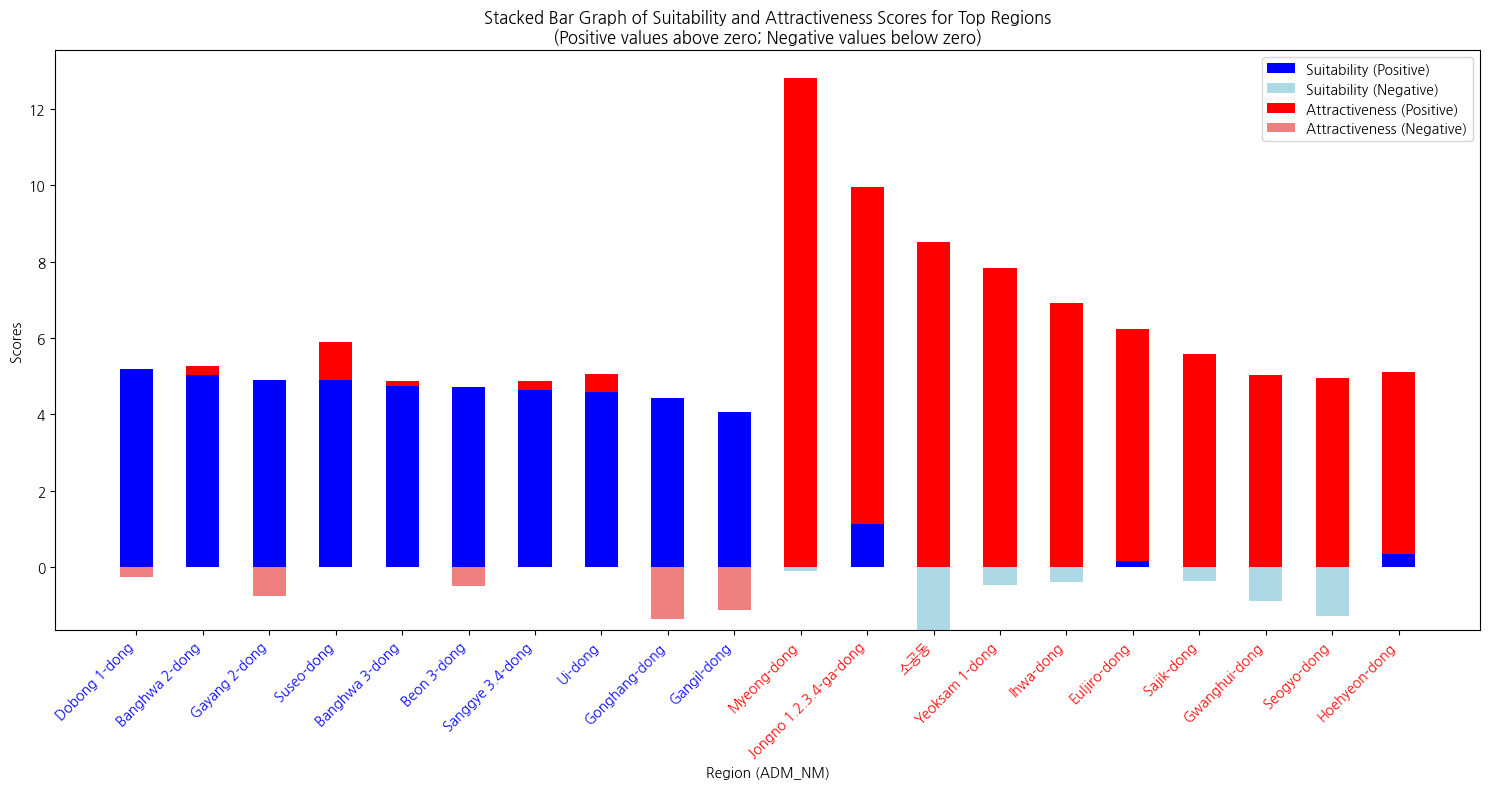}
  \caption{Top 10 Regions for Suitability (blue-colored) and Attractiveness Scores (red-colored)}\label{fig:stacked_SA}
\end{figure}

Figure \ref{fig:stacked_SA} highlights the contrasting characteristics of Seoul’s top 10 regions for suitability and attractiveness. Suitability leaders demonstrate uniform score distributions from Dobong 1-dong (Suitability Score: 5.19) to Ganggil-dong (4.06) making all regions stable candidates for phased UAM deployment. Conversely, attractiveness scores show extreme spatial skewness, dominated by central hubs like Myeong-dong (Attractiveness Score: 12.81) and Jongno 1.2.3.4-dong (8.81). This divergence implies the distinct implementation strategies: concentrated initial deployments in high-attractiveness zones versus distributed networks prioritizing suitability.

\subsection{Factor Decomposition of Top-Scoring Regions}
\begin{figure}[!ht]
  \centering
  \includegraphics[width=1\textwidth]{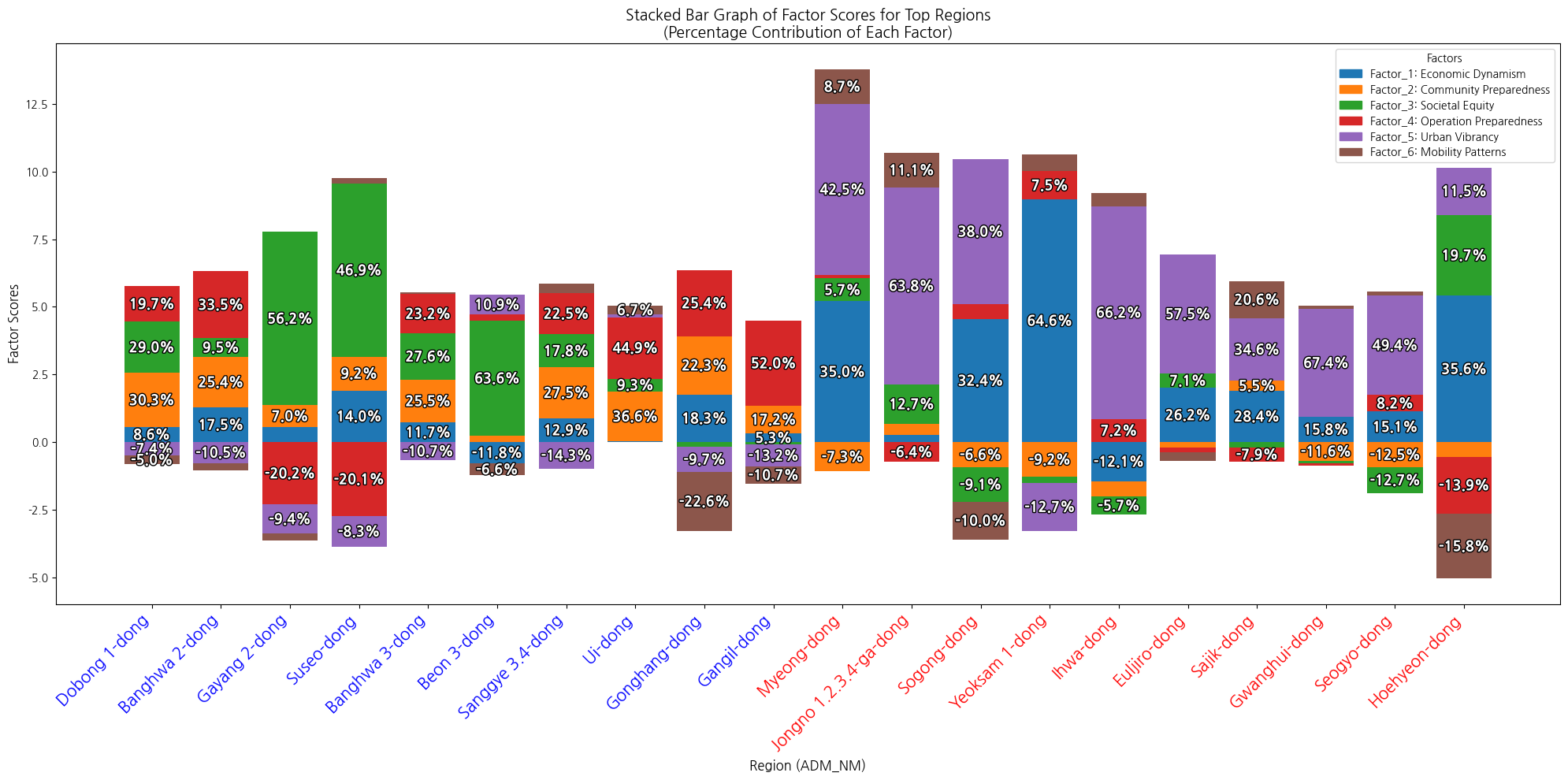}
  \caption{Factor scores for the top 10 suitability and attractiveness regions with absolute contribution in percentages}\label{fig:stacked}
\end{figure}

Figure \ref{fig:stacked} reveals the latent factor compositions driving regional scores. When percentage contributions are examined, the high-suitability regions (left cluster) derive 46.7\% of their scores from Factor 2 (Community Preparedness) and Factor 3 (Societal Equity) on average, emphasizing dominant contribution of these latent factors on suitability. In contrast, top-attractiveness regions (right cluster) rely predominantly (63.3\%) on Factor 5 (Urban Vibrancy) and Factor 1 (Economic Dynamism), with Yeoksam-1-dong showing exceptional Factor 1 dominance (64.6\%).

The figure highlights that suitability leaders exhibit significant positive contributions from Factor 2 (mean 20.3\%), emphasizing its critical role in determining regional suitability. In contrast, Factor 2 scores are mostly negative or negligible in regions with high attractiveness. Conversely, Factor 5 demonstrates the opposite trend, further underscoring key differences between suitability and attractiveness. These findings suggest that community preparedness and urban vibrancy within the urban environment act as pivotal discriminators between suitability and attractiveness.

Insights derived from Seoul's analysis can be extended to scenarios where regional data is scarce or difficult to obtain. Without the need to implement the full USE-LFA framework, decision-makers in cities with urban environments similar to Seoul can leverage community preparedness and urban vibrancy as proxies to approximate regional suitability and attractiveness. This demonstrates the potential applicability of the USE-LFA framework in cities with limited data availability.

\subsection{Quadrant Map of Suitability and Attractiveness}
\begin{figure}[!ht]
  \centering
  \includegraphics[width=0.9\textwidth]{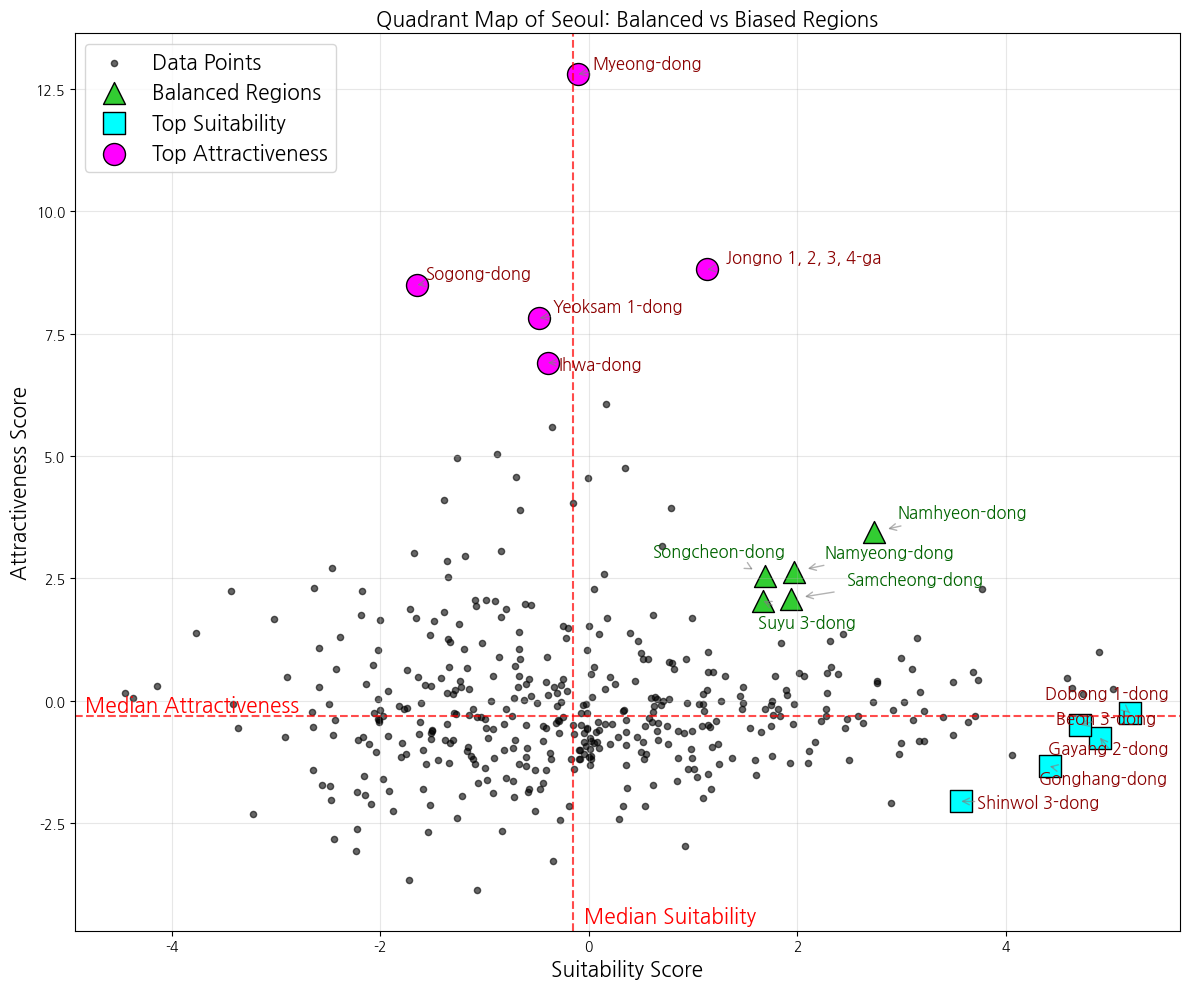}
  \caption{Quadrant Map of Suitability and Attractiveness Scores in Seoul (Magenta: Attractiveness-biased; Cyan: Suitability-biased; Green: Balanced Regions)}\label{fig:quadrant}
\end{figure}

The quadrant map in Figure \ref{fig:quadrant} visualizes the spatial distribution of regions in Seoul based on their suitability and attractiveness scores. By defining median thresholds for both metrics, the map categorizes regions into four quadrants, revealing distinct spatial typologies and clustering patterns.

Three primary typologies emerge. Balanced regions shown in green triangles represent areas with high scores for both suitability and attractiveness, with minimal differences between the two scores. Examples include Namhyeon-dong, Namyeong-dong, and Songcheon-dong. These regions are likely to offer comprehensive benefits due to their equilibrium between urban qualities. Suitability-biased regions in cyan squares exhibit the largest score differences, with suitability scores dominating. Examples include Dobong 1-dong, Beon 3-dong, and Gayang 2-dong. Conversely, attractiveness-biased regions drawn in magenta circles, such as Myeong-dong, Sogong-dong, and Yeoksam 1-dong, demonstrate exceptionally high attractiveness scores with significant differences compared to their suitability scores.

The broader distribution of data points reveals several notable patterns. (1) Most regions cluster near median values, indicating moderate performance in both metrics. (2) An elliptical cloud stretching from lower left to upper right suggests a loose positive correlation between suitability and attractiveness. (3) The lower left quadrant highlights under-performing regions that may require targeted urban renewal efforts. (4) High-density regions near the origin (median intersection) represent typical Seoul regions with average urban qualities. These typologies highlight Seoul's urban configuration while offering valuable insights for policy applications. Cities with similar quadrant distributions could adopt transferable strategies tailored to their specific typologies.

An intriguing observation is that regions with extreme biases in either suitability or attractiveness are not positioned near the lowest values of the opposite metric. Instead, they tend to cluster around the median point of the opposite scoring group. This suggests that regions with pronounced strengths in one dimension still maintain moderate performance in the other dimension rather than being entirely deficient.

While the quadrant map provides intuitive visual insights into regional typologies, its reliance on subjective thresholds may limit its applicability in scenarios requiring adaptable prioritization between suitability and attractiveness. To address this limitation and enhance decision-making flexibility, we introduce the concept of the v-score.

\subsection{\textit{v-score}}
The v-score is a composite metric that combines suitability and attractiveness scores using a weighting parameter $\alpha$. This parameter, which ranges from 0 to 1, allows users to adjust the relative importance of suitability and attractiveness based on preference states or strategy-specific priorities. The formulation of the v-score confines suitability and attractiveness scores as follows:
\begin{equation}
    \textit{v-score} = \alpha(suitability\_score) + (1-\alpha)(attractiveness\_score).
    \label{eq:v-score}
\end{equation}

\subsubsection{Spatial Sensitivity to Different Weighting Parameter}
\begin{figure}[!ht]
  \centering
  \includegraphics[width=1\textwidth]{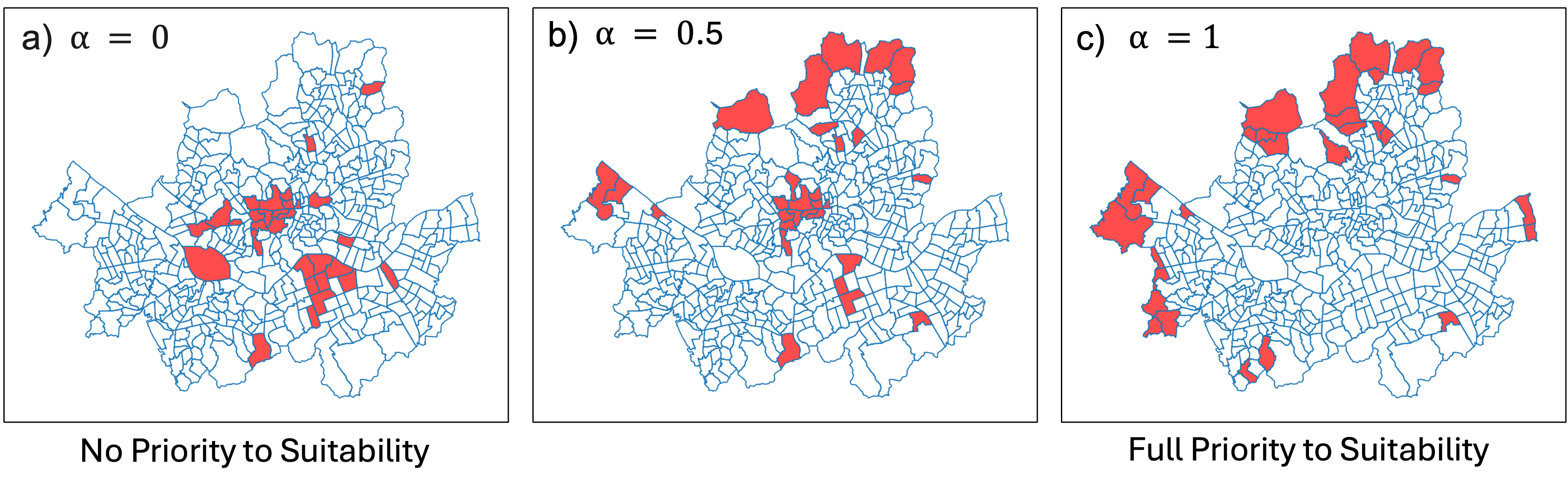}
  \caption{Spatial distribution of the top 30 regions with the highest v-scores under varying levels of suitability prioritization (\(\alpha\)).}\label{fig:prioritization_3}
\end{figure}

When implementing different prioritization parameters, clear spatial variations emerge. Figure \ref{fig:prioritization_3} illustrates how the distribution of high-scoring regions shifts as $\alpha$ increases from 0 (full attractiveness priority) to 1 (full suitability priority).

At \(\alpha=0\), where attractiveness is fully prioritized, high v-score regions cluster in affluent, economically prominent areas of Seoul, reflecting the strong link between attractiveness and established economic centers. As \(\alpha\) increases to 0.5, representing an equal prioritization of suitability and attractiveness, the distribution shifts, with high v-score regions becoming more dispersed and extending towards the outer parts of Seoul. At \(\alpha=1\), fully prioritizing suitability, high v-score clusters reappear, primarily in the norther and western parts of Seoul. These transitions demonstrate how prioritization significantly influences optimal site selection, suggesting that balanced approaches may identify different locations than those focusing solely on market potential or operational requirements. To further analyze the impact of different prioritization, the convex combinations were examined with the contour plots of the v-scores.

\subsubsection{Convex Combination of Suitability and Attractiveness}
\begin{figure}[!ht]
  \centering
  \includegraphics[width=1\textwidth]{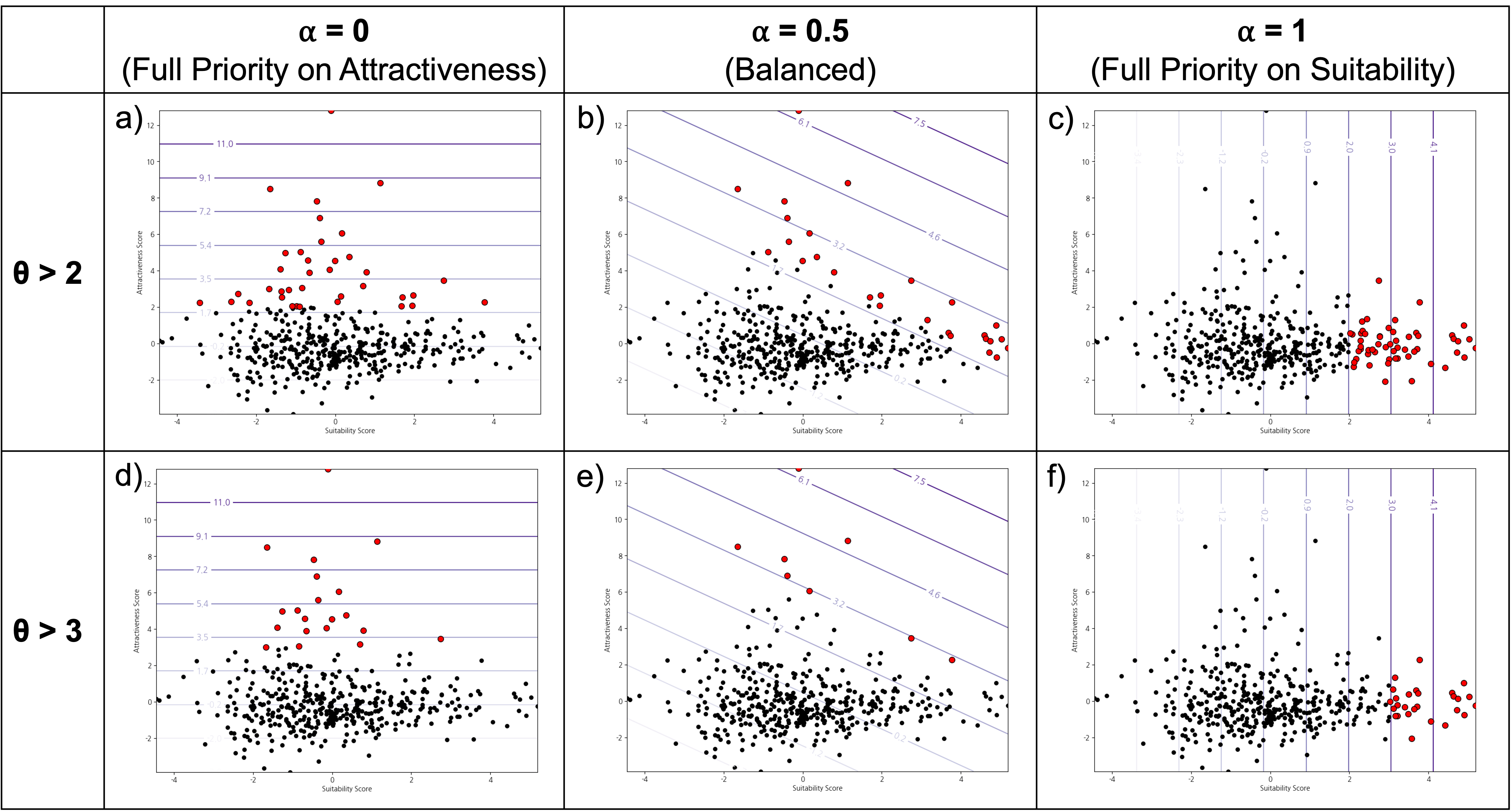}
  \caption{Selected regions in red dots with the v-score higher than the thresholds 2 (from a to c) and 3 (from d to f)}\label{fig:convex_comb}
\end{figure}

The analysis of the convex combination of suitability and attractiveness scores reveals significant insights into potential vertiport site selection. Three different weights were examined, with the slope of the v-score contours varying as the weight for suitability increased. This demonstrates the growing influence of suitability on the composite v-score as its weight increases. When applying a v-score threshold of $\theta > 2$, the number of identified regions changes notably. At weight 0 (37 selected regions), significantly fewer regions meet the criteria than at weight 1 (55 regions). This disparity highlights the skewed nature of attractiveness scores and suggests that fewer regions excel in attractiveness compared to suitability.

\subsubsection{Threshold-Dependent Weighting Effects}
\begin{figure}[!ht]
  \centering
  \includegraphics[width=0.9\textwidth]{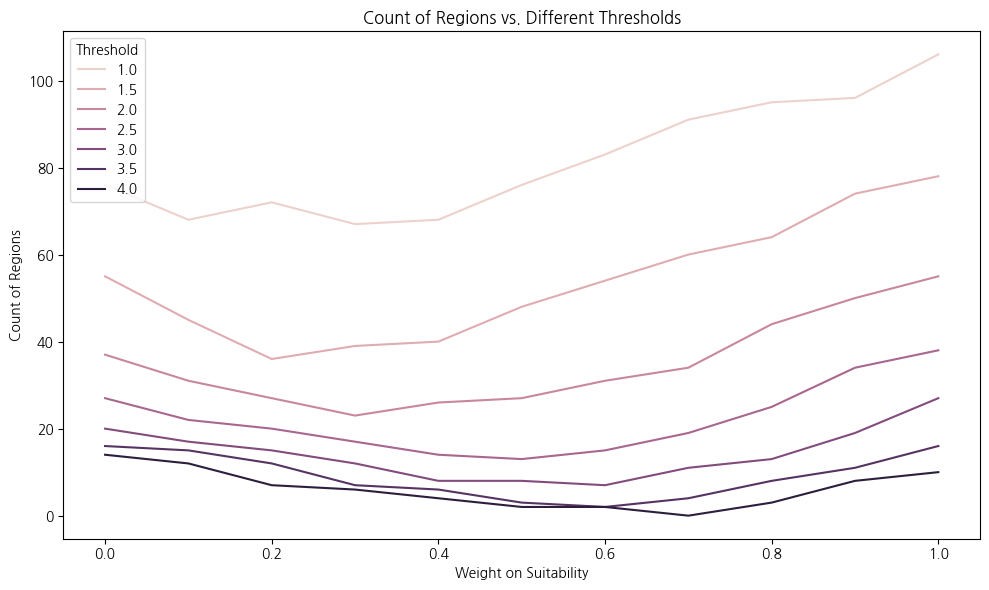}
  \caption{illustrates the relationship between preference weight $\alpha$ and the count of regions across seven different $\theta$ values.}\label{fig:lineplot}
\end{figure}

Analysis of scoring thresholds and weights revealed distinct patterns in the number of candidate regions. Figure \ref{fig:lineplot} shows the relationship between preference weight ($\alpha$) and region counts across seven threshold values ($\theta$). All curves exhibit a convex shape: region counts initially decrease as $\alpha$ increases from 0, reach a minimum at intermediate $\alpha$ values (typically 0.2-0.6), then increase as $\alpha$ approaches 1.0. This pattern indicates that balanced weighting creates the most stringent filtering effect, suggesting regions excelling in both dimensions are less common than those performing well in just one.

Table \ref{tab:lambda_threshold_counts} quantifies these relationships, showing region counts and percentages out of 426 total regions in Seoul for different threshold-weight combinations. Sensitivity to weighting is more pronounced at lower thresholds ($\theta \leq 1.5$), where the changes curves are steeper and the percentage difference between the maximum and minimum number is greater. At higher thresholds, this sensitivity diminishes. The minimum points on most curves occur more during the lower values of the weight (around $\alpha$ = 0.4), indicating suitability is somewhat more stringent measure than attractiveness. 

\begin{table}[ht]
\centering
\small
\begin{tabular}{c|cccccc}
\hline
Threshold ($\theta$) & \multicolumn{6}{c}{Weight ($\alpha$)} \\
\cline{2-7}
& 0.0 & 0.2 & 0.4 & 0.6 & 0.8 & 1.0 \\
\hline
1.0 & 76 (17.8\%) & 72 (16.9\%) & 68 (16.0\%) & 83 (19.5\%) & 95 (22.3\%) & 106 (24.9\%) \\
1.5 & 55 (12.9\%) & 36 (8.5\%) & 40 (9.4\%) & 54 (12.7\%) & 64 (15.0\%) & 78 (18.3\%) \\
2.0 & 37 (8.7\%) & 27 (6.3\%) & 26 (6.1\%) & 31 (7.3\%) & 44 (10.3\%) & 55 (12.9\%) \\
2.5 & 27 (6.3\%) & 20 (4.7\%) & 14 (3.3\%) & 15 (3.5\%) & 25 (5.9\%) & 38 (8.9\%) \\
3.0 & 20 (4.7\%) & 15 (3.5\%) & 8 (1.9\%) & 7 (1.6\%) & 13 (3.1\%) & 27 (6.3\%) \\
3.5 & 16 (3.8\%) & 12 (2.8\%) & 6 (1.4\%) & 2 (0.5\%) & 8 (1.9\%) & 16 (3.8\%) \\
4.0 & 14 (3.3\%) & 7 (1.6\%) & 4 (0.9\%) & 2 (0.5\%) & 3 (0.7\%) & 10 (2.3\%) \\
\hline
\end{tabular}
\caption{Counts and percentages (out of 426 regions), step increment of $\alpha = 0.2$.}
\label{tab:lambda_threshold_counts}
\end{table}

These findings demonstrate the critical role of strategic weighting decisions in reshaping candidate region selection. The observed convex patterns and threshold sensitivity highlight the necessity for stakeholders to explicitly prioritize either suitability or attractiveness during site selection processes. Balanced weighting ($\alpha = 0.4-0.6$) emerges as a stringent filter, favoring regions that excel in both dimensions, while extreme weight values emphasize specialization in one dimension over the other. This insight provides a foundation for informed decision-making, ensuring that site selection strategies align with strategic objectives.

\section{Conclusion}
This study presents the Urban Site Evaluation with Latent Factor Analysis (USE-LFA) framework, a comprehensive, data-driven methodology for evaluating urban port sites, demonstrated through a case study of vertiport site selection in Seoul. By integrating Latent Factor Analysis (LFA), the framework identifies six key urban factors-Economic Dynamism, Community Preparedness, Societal Equity, Operational Preparedness, Urban Vibrancy, and Mobility Patterns-categorized into two dimensions: \textit{Suitability} and \textit{Attractiveness}. These enable a dual-perspective evaluation of urban environments, balancing operational feasibility and operational merit. The framework’s adaptability is underscored by the development of a composite \textit{v-score}, which allows stakeholders to prioritize suitability or attractiveness based on strategic objectives. The analysis revealed how optimal site selection shifts spatially depending on prioritization, highlighting the tension between centralized economic hubs and more dispersed operationally viable regions. The quadrant-based classification of regions and convex combination analysis further provide informative insights for tailoring urban port integration strategies. 

Unlike existing tools such as NASA’s VAMOS! \citep{sheth2023vamos}, which often conflate operational feasibility and merits into a single evaluation, USE-LFA distinctly separates these dimensions while integrating broader urban environmental factors such as community preparedness and societal equity. This separation enhances the framework’s versatility, allowing it to adapt to diverse urban contexts and enabling stakeholders to evaluate suitability and attractiveness independently while tailoring decisions to specific regional needs. Furthermore, its flexible design supports applications beyond vertiports, such as transit hubs and other emerging mobility infrastructures. By systematically addressing the interplay between operational feasibility and merits, USE-LFA provides informative insights for the systematic integration of ports into complex urban environments. Policymakers can leverage this framework to align port placement with strategic goals such as improving accessibility in under-served areas or enhancing emergency response capabilities. Urban planners can use factor-specific insights to design infrastructure that complements existing systems while addressing community concerns. Moreover, the framework’s flexibility allows for continuous adaptation to evolving urban environments or new data inputs.

Future research can expand our framework by incorporating additional layers such as noise impact assessments \citep{tan2023virtual, tan2024exploring}, airspace constraints, and dynamic urban changes. For example, integrating optimized noise-aware flight trajectories with regional characteristics could support sustainable UAM adoption \citep{ho2019air, gao2023developing}. By further addressing these components, our framework could be utilized to ensure that UAM systems remain equitable and efficient across diverse contexts. 

\hypertarget{acknowledgements}{%
\section*{Acknowledgements}}
This work is financially supported by Korea Ministry of Land, Infrastructure, and Transport (MOLIT) as \textit{Innovative Talent Education Program for Smart City} and by the National Research Foundation of Korea(NRF) grant funded by the Korea government (MSIT) (No. 2020R1A2C2010200).

\hypertarget{GenerativeAI}{%
\section*{Generative AI and AI-assisted Technologies}}
During the preparation of this manuscript, the authors used ChatGPT and Grammarly to check the grammar. After using this tool/service, the authors reviewed and edited the content as needed and take full responsibility for the content of the publication.

\hypertarget{contributions}{%
\section*{Authors Contributions}}
\textbf{Sungmin Sohn}: Conceptualization, Methodology, Software, Formal Analysis, Investigation, Data Curation, Writing – Original Draft, Visualization, Project administration. \textbf{Namwoo Kim}: Conceptualization, Methodology, Validation, Resources, Supervision. \textbf{Mark Hansen}: Validation, Investigation, Supervision. \textbf{Yoonjin Yoon}: Conceptualization, Methodology, Validation, Resources, Writing – Review \& Editing, Supervision, Project administration, Funding acquisition.

\selectlanguage{english}

\bibliographystyle{plainnat}
\bibliography{main}

\appendix

\section{Notation Table with Descriptions}
The following notation establishes the fundamental components of our urban latent factor analysis for vertiport site selection (Table \ref{tab:notation}).

\begin{table}[htbp]
\centering

\caption{Definition of Notation for Latent Factor Analysis}
\label{tab:notation}
\begin{tabular}{@{}>{\raggedright}p{2.3cm}>{\raggedright}p{2.7cm}>{\raggedright\arraybackslash}p{9cm}@{}}
\toprule
\textbf{Symbol} & \textbf{Dimensions} & \textbf{Description} \\ 
\midrule
$R$ & $\mathbb{N}$ & Number of regions in the study area. \\ 
$N$ & $\mathbb{N}$ & Number of urban attributes measured. \\ 
$M$ & $\mathbb{N}$ & Number of latent factors extracted. \\ 
\midrule
$\mathbf{A}=[a_{ij}]$ & $\mathbb{R}^{N\times R}$ & Standardized urban attribute matrix; $a_{ij}$ is the standardized measure of attribute $i$ in region $j$. \\ 
$\mathbf{L}=[\ell_{im}]$ & $\mathbb{R}^{N\times M}$ & Urban attribute loading matrix; $\ell_{im}$ represents the intensity of relationship between attribute $i$ and latent factor $m$. \\ 
$\mathbf{F}=[f_{mj}]$ & $\mathbb{R}^{R\times M}$ & Urban factor score matrix; $f_{mj}$ indicates the degree to which region $j$ embodies latent factor $m$. \\ 
$\mathbf{E}$ & $\mathbb{R}^{N\times R}$ & Error matrix capturing region-specific variance unique to each attribute. \\ 
\midrule
$\mathbf{R}$ & $\mathbb{R}^{N\times N}$ & Correlation matrix of raw urban attributes. \\ 
$\hat{\mathbf{c}}_i^{2(0)}$ & $\mathbb{R}^{+}$ & Initial communality for attribute $i$, calculated as $1-\frac{1}{r^{ii}}$ where $r^{ii}$ is the $i^\text{th}$ diagonal element of $\mathbf{R}^{-1}$. \\ 
$\mathbf{R}^*$ & $\mathbb{R}^{N\times N}$ & Adjusted correlation matrix with diagonal elements replaced by communality estimates. \\ 
\midrule
$\mathbf{Q}$ & $\mathbb{R}^{N\times N}$ & Eigenvector matrix from decomposition of $\mathbf{R}^*$; columns are eigenvectors $\boldsymbol{q}_m$. \\ 
$\mathbf{\Lambda}$ & $\mathbb{R}^{N\times N}$ & Diagonal matrix of eigenvalues $\lambda_1,...,\lambda_N$ from decomposition of $\mathbf{R}^*$. \\ 
$\mathbf{L}^{(k)}$ & $\mathbb{R}^{N\times M}$ & Factor loading matrix at iteration $k$, computed from eigenvectors and eigenvalues. \\ 
\midrule
$\hat{\mathbf{c}}_i^{2(k+1)}$ & $\mathbb{R}^{+}$ & Updated communality for attribute $i$ at iteration $k+1$, calculated as $\sum_{m=1}^M (\ell_{im}^{(k)})^2$. \\ 
$\epsilon$ & $\mathbb{R}^{+}$ & Convergence threshold ($10^{-5}$) for terminating iterative updates. \\ 
$\mathbf{L}^*$ & $\mathbb{R}^{N\times M}$ & Final factor loading matrix containing $M$ factors that satisfy Kaiser criterion ($\lambda_m \geq 1$). \\
\midrule
$\mathbf{L}^R$ & $\mathbb{R}^{N\times M}$ & Rotated factor loading matrix after Varimax rotation for improved interpretability. \\
$\mathbf{T}$ & $\mathbb{R}^{M\times M}$ & Orthogonal rotation matrix satisfying $\mathbf{T}^\top\mathbf{T}=\mathbf{I}_M$. \\ 
$\mathbf{B}=[\beta_{mi}]$ & $\mathbb{R}^{M\times N}$ & Factor scoring weights calculated as $(\mathbf{L}^\top\mathbf{R}^{-1}\mathbf{L})^{-1}\mathbf{L}^\top\mathbf{R}^{-1}$. \\ 
\midrule
$\mathbf{T}$ & $\{0,1\}^{2\times M}$ & Composite definition matrix; $t_{1m}=1$ if factor $m$ belongs to Suitability, $t_{2m}=1$ if Attractiveness. \\ 
$\hat{\mathbf{F}}$ & $\mathbb{R}^{2\times R}$ & Composite score matrix computed as $\mathbf{T}\mathbf{F}$, containing Suitability and Attractiveness scores for each region. \\
$\mathcal{S}$ & $\subset\{1,...,M\}$ & Set of indices for Suitability factors. \\
$\mathcal{A}$ & $\subset\{1,...,M\}$ & Set of indices for Attractiveness factors, where $\mathcal{S}\cap\mathcal{A}=\emptyset$.\\
\bottomrule
    \end{tabular}
\end{table}

\end{document}